\documentstyle[aps,prl,epsfig]{revtex}

\begin{document}
\draft
\twocolumn[\hsize\textwidth\columnwidth\hsize\csname %
@twocolumnfalse\endcsname

\title{Vanishing Hall Constant in the Stripe Phase of Cuprates}
\author{P.\ Prelov\v sek$^1$, T.\ Tohyama$^2$, and S.\ Maekawa$^2$}
\address{$^1$ Faculty of Mathematics and
Physics, University of Ljubljana, and J.\ Stefan Institute, 1000
Ljubljana, Slovenia }
\address{$^2$ Institute for Materials Research, Tohoku University,
Sendai 980-8577, Japan}
\date{\today}
\maketitle
\begin{abstract}\widetext
The Hall constant $R_H$ is considered for the stripe structures. In
order to explain the vanishing of $R_H$ in
La$_{2-y-x}$Nd$_{y}$Sr$_x$CuO$_4$ at $x=1/8$, we use the relation of
$R_H$ to the Drude weight $D$ as well as direct numerical calculation,
to obtain results within the $t$-$J$ model, where the stripes are
imposed via a charge potential and a staggered magnetic field. The
origin of $R_H\sim 0$ is related to a maximum in $D$ and the minimal
kinetic energy in stripes with a hole filling $\sim 1/2$. The same
argument indicates on a possibility of $R_H\sim 0$ in the whole range
of static stripes for $x \leq 1/8$.

\end{abstract}
]
\narrowtext
The evidence for the stripe structures in
La$_{2-y-x}$Nd$_{y}$Sr$_x$CuO$_4$ (LNSCO) emerging from neutron
scattering experiments \cite{tran} has stimulated in recent years
numerous experimental and theoretical investigations, trying to relate
this phenomenon to superconductivity and other anomalous properties of
cuprates. In the commensurate case $x \sim 1/8$ the stripe structure
represents spin and charge ordering, i.e., domain walls within an
ordered two-dimensional antiferromagnet (AFM) with the filling of
$n_l=N_h/L=1/2$ hole per unit length (quarter filling in the usual
band picture) within each domain wall - stripe. Actually, it was shown
in the recent angle-resolved photoemission spectroscopy experiments
that the stripes are in the quarter filled state \cite{zhou}. In
contrast to many other systems showing spin-density wave and
charge-density wave order at low temperatures, the stripe structures
in cuprates appear to be metallic \cite{uchi1}. This can be explained
with charge carriers - holes being well mobile along the
one-dimensional (1D) stripes. Recently, a systematic study of LNSCO
($y=0.6$) with varying doping $x$
\cite{uchi2} revealed a striking signature of stripes in the behavior
of the Hall constant $R_H$. While for $x>0.15$ $R_H(x,T)$ is very
close to that of the reference doped La$_{2-x}$Sr$_x$CuO$_4$ (LSCO),
$R_H$ appears to be very sensitive to the structural phase transition
between the low-temperature tetragonal (LTT) and orthorhombic (LTO)
phase for $x<0.15$. Below the LTT-LTO transition $R_H$ becomes
strongly $T$ and $x$ dependent and it eventually vanishes, $R_H \sim
0$, at $T\sim 0$ for $x\sim 0.12$, i.e., in the structure with long
range stripe order. At the same time the anomaly is weakly pronounced
in the planar resistivity $\rho_{ab}$ \cite{uchi2}.

In the present paper, we investigate $R_H$ in stripe structures and in
particular a possible explanation for its vanishing.  $R_H \sim 0$ in a
conducting system is quite  unusual and non universal. For instance, it
could happen in an ordinary metal due to (accidental) cancellation of
band curvatures or due to band crossings. However, cuprates are closer
to hole-doped Mott insulators where in the reference LSCO a
(semiconductor-like) semi-classical $R_H
\propto 1/x$ is followed in a wide range $x<0.2$ at low $T$. $R_H \sim 0$ in
LNSCO at $x \sim 1/8$ is, on the other hand, a very pronounced
deviation from the latter. Therefore, one can speculate on more
fundamental origin of this most pronounced macroscopic effect of
stripes whereby the strong correlations play an essential role. The
authors \cite{uchi2} interpreted the experimental finding $R_H \sim 0$
as an evidence for a transport along independent stripes. This
requires exclusively 1D transport and therefore can hardly apply.
Hall constant can be expressed in terms of planar conductivities
$\sigma_{\alpha\beta}$ in a finite magnetic field $B$ as
$R_H=\sigma_{xy}/[B\sigma_{xx}\sigma_{yy}]$. Assuming a very
anisotropic system with stripes, e.g., along the $x$ direction and large
$\sigma_{xx}$, it is plausible that $\sigma_{xy}$ scales with the weak
hopping between stripes. But $\sigma_{yy}$ is expected usually to
scale in the same way leaving the ratio $R_H$ finite. The evidence for
the latter behavior can be found in the experiments on very
anisotropic quasi-1D metals which reveal nevertheless quite normal
metallic values of $R_H$ \cite{coop}.

Another simple interpretation is that in a stripe structure with hole
filling $n_l=1/2$ we are dealing along the single stripe with an equal
concentration of holes and electrons, leading to the cancellation of
$\sigma_{xy}$ while both diagonal $\sigma_{\alpha\alpha}$ remain
finite. Although it is hard to improve this argument into a more
formal theory, to some extent it seems to be closer to our final
conclusions further on.

It is evident that the Hall effect in cuprates \cite{ong,hwan}, as
well as in systems with strongly correlated electrons in general,
still lacks proper theoretical understanding. This applies for most
investigated reference cuprate LSCO and $R_H$ dependence on $T$ and
hole doping $n_h$. While the theoretical formulation of the Hall
linear response $R_H(\omega,T)$ is well established \cite{fuku,shas},
there are very few theoretical results and numerical studies for
models with strong correlation in particular in the relevant low $T$
and low doping regime \cite{cast}.  Recently an approach and a
numerical procedure have been designed \cite{prel1} which yield for a
static d.c. $R_H$ at $T=0$ (although for a ladder system) the desired
result, namely a semi-classical behavior $R_H^*=1/e_0 n_h$ in a
magnetic insulator at low hole doping $n_h=N_h/N$ ($e_0$ being 
the electric charge), as found
experimentally in cuprates \cite{ong,hwan}. Moreover, for a reactive
(non-dissipative) $R_H(T=0)$ \cite{zoto} it was possible to find an
useful relation to the variation of the Drude weight (charge
stiffness) $D$ with the electron density $n$,
\begin{equation}
R_H= - \frac{1}{e_0 D}\frac{\partial D}{\partial n}. \label{eq1}
\end{equation}
The latter relation is derived using the non-standard limit: $\omega
\to 0$ first, then $q \to 0$. Nevertheless, it has very attractive
properties for the analysis of strongly correlated electrons: a) $D$
at $T=0$ is the central quantity distinguishing the Mott-Hubbard
insulator from a conductor, b) close to the Mott insulator $D \propto
n_h=1-n$ directly implies a plausible semi-classical result
$R_H^*=1/e_0 n_h$, which is hard to be established by other methods.
While so far the relation has not been justified for the general case
in the proper transport limit ($q \to 0$ first, then $\omega \to 0$)
it appears to be valid for a strongly anisotropic system where
$D=D_{\alpha\alpha}$ should be taken along the most conducting
direction $\alpha$
\cite{prel2}. 

There is also a qualitative similarity of the relation (\ref{eq1})
with the proposal in Ref.\cite{rojo} where the authors correlate Hall
($B>0$) conductivity $\sigma_{xy}$ with the doping dependence of the
kinetic energy, i.e., $\partial \langle K\rangle/\partial n$. Note
that in a (tight binding) strongly correlated system $K$ and $D$ are
related through the sum rule
\begin{equation}
N D_{\alpha\alpha}= \frac{1}{2}\langle -K_{\alpha\alpha} \rangle-
\frac{1}{e_0^2} \int_0^\infty
\sigma_{\alpha\alpha}^{reg}(\omega) d\omega,
\label{eq2}
\end{equation}
where $\sigma^{reg}$ is the regular part of the optical conductivity
and $K_{\alpha\alpha}$ involves only the hopping in the $\alpha$
direction.

Theoretical modeling of stripes in cuprates has proven to be quite
delicate just due to their $n_l=1/2$ filling and metallic
character. Stripe solutions of relevant (Hubbard and analogous) models
within simplest mean-field approximations generally show a tendency
towards $n_l=1$ filling and an insulating behavior. More recently
extensive numerical analyses within the relevant $t$-$J$ model
\cite{whit} confirm the relative stability of metallic stripes at
commensurate doping $n_h=1/8$. These results and other approaches
\cite{prel0,tohy,cher} all indicate a crucial interplay of the 
hole kinetic energy and the spin Heisenberg interaction in
metallic stripes, both being crucially related to strong electron
correlation.

Equation (\ref{eq1}) suggests an explanation of $R_H \sim 0$ in
$x=1/8$ stripes, namely that $D(n)$ is a local maximum for $n_l=1/2$,
this being closely related to the minimum kinetic energy $\langle
K\rangle$ in this configuration.  In order to support this conjecture
we perform a numerically exact-diagonalization study of $D$ and $R_H$
within the prototype $t$-$J$ model,
\begin{equation}
H=-t\sum_{\langle ij\rangle  s}(\tilde{c}^\dagger_{js}\tilde{c}_{is}+
{\rm H.c.})+J\sum_{\langle ij\rangle} ({\bf S}_i\cdot {\bf S}_j -
{1\over 4} n_i n_j), \label{eq3}
\end{equation} 
where no double occupancy of sites is allowed. In numerical studies we
use $J/t=0.4$ to be in the strong-correlation regime relevant to
cuprates. 

Since studied systems are too small to inherently reproduce stripes,
we impose them through an attractive hole potential $V$ along the
stripe \cite{tohy} and the staggered field $h_i=\pm h$ forming an AFM
phase boundary across the stripe \cite{prel0,cher},
\begin{equation}
H_{st}=-V\sum_{i \in st} (1-n_i) + \sum_{i \notin st} h_i S^z_i.
\label{eq4}
\end{equation}

\begin{figure}[t]
\begin{center}
\epsfig{file=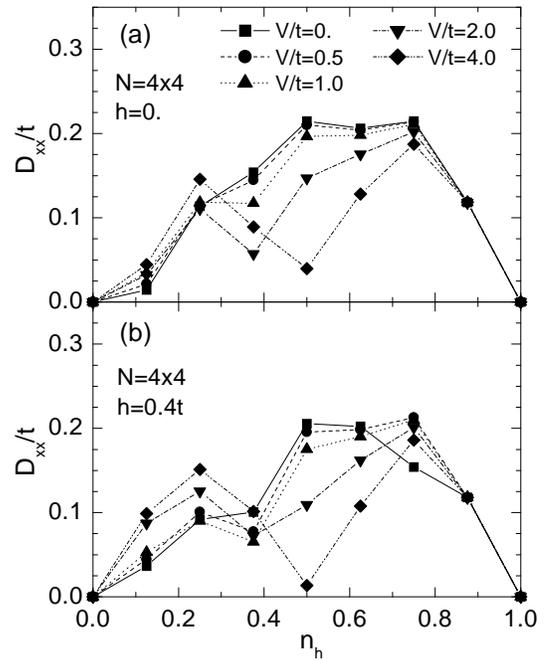,width=70mm,clip=}
\end{center}
\caption{Drude weight $D_{xx}/t$ as a function of doping $n_h$ calculated
for a $4\times 4$ system with p.b.c., where the stripes are imposed in
the $x$ direction, for various stripe potentials $V$ and staggered
fields: (a) $h=0$ and (b) $h=0.4~t$.}
\label{fig1}
\end{figure}

First we consider numerically a system with $N=4\times 4$ sites and
periodic boundary conditions (p.b.c.) in both the $x$ and $y$ directions.
Two stripes are imposed along the $x$-direction with the inter-stripe
distance $d_0=2$. $D_{xx}=D$ in such a system can be evaluated by
introducing a flux through the torus modifying $t \to
\tilde t_{ij} = t~{\rm exp}[i \theta (x_i -x_j)]$
to find an absolute energy minimum $E^0(\theta)$ at $\theta=\theta_0$
and then by taking the second derivative, i.e., $D=(1/2N) \partial^2
E^0(\theta)/\partial \theta^2|_{\theta=\theta_0}$.  In Figs.~1(a) and
(b), we present results for $D$ as a function of hole doping $n_h$
(only even $N_h$ are considered) and various stripe potentials $V$,
without the staggered field $h=0$ as well as with $h/t=0.4$.  Two
limits for $D(n_h)$ are quite easy to understand. For $V=0$ we are
dealing with a homogeneous 2D $t$-$J$ model where $D$ reaches maximum
at $n_h \sim 0.5$.  On the other hand, for large $V\gg t$ holes are
localized to a 1D motion within each stripe. As in 1D, $D$ then
reaches the maximum for stripe filling $n_l=1/2$, i.e., for $n_h=1/(2
d_0)=0.25$ (while $D \sim 0$ for $n_l \sim 1$).  Using Eq.(\ref{eq1})
in the latter case clearly leads to $R_H \sim 0$ for stripes at
$n_l=1/2$. Our results in Fig.~1(a) suggest that the maximum remains
pronounced also for far more modest $V\geq t$. The introduction of the
staggered field $h$, as shown in Fig.~1(b), enhances but also flattens
the maximum which is visible now even at $V/t \sim 0.5$.

\begin{figure}[t]
\begin{center}
\epsfig{file=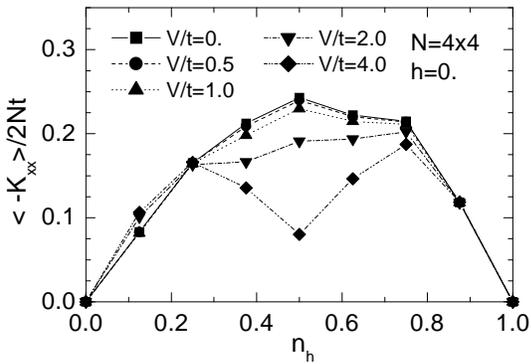,width=70mm,clip=}
\end{center}
\caption{Kinetic energy $\langle -K_{xx}\rangle /2Nt$ vs. $n_h$ for
a $4\times 4$ system with various $V/t$ at $h=0$.}
\label{fig2}
\end{figure}

For comparison we present in Fig.~2 the related variation of $\langle
-K_{xx}\rangle$ with $n_h$. We notice that in analogy to $D$ there is
a maximum at $n_l \sim 1/2$, however well pronounced only for larger
$V/t$ whereby $\langle K_{xx}\rangle$ is also less sensitive to $h$.

By considering the $t$-$J$ model with open boundary conditions
(o.b.c.) in the $y$ direction, i.e., on a ladder system $N=L \times
M$, one can study $R_H$ even more directly \cite{prel1}. If we
introduce the Hall electric field in the $y$-direction as
$H_\Delta=\Delta \sum_i (y_i-
\bar y) n_i$ as well as the magnetic field $B=\varphi/e_0$
perpendicular to the planar system entering the hopping $\tilde t_{ij}
\to \tilde t_{ij} {\rm exp}[i \varphi y_i (x_i -x_j)]$, we can express
$R_H$ in terms of derivatives of the ground state
$E^0(\theta,\Delta,\varphi)$ with respect to external fields
\cite{prel1},
\begin{equation}
R_H= - \frac{ N E^0_{\theta\Delta\varphi}}{e_0 E^0_{\Delta
\Delta} E^0_{\theta \theta}}\Bigl|_{\Delta=\phi=0,\theta =\theta_0},
\label{eq5}
\end{equation}
where derivatives should be taken in general at $\theta =\theta_0$,
where $E^0$ is minimum.

For a reasonable application of Eq.(\ref{eq5}), a system with $L>M$ is
required. We therefore restrict our numerical calculations to a
$6\times 3$ system with o.b.c. in the M direction, where a single
stripe is introduced via $H_{st}$, Eq.(\ref{eq4}), in the middle
leg. The situation with stripe filling $n_l=1/2$ here corresponds to
$N_h=3$ holes in the system.  Odd $N_h$ has some disadvantage leading
in general to $\theta_0\neq 0$ and also to near-degenerate states at
$V=0$. As a reference for $V=0$ one should expect at low doping $n_h
\ll 1/2$ the semi-classical behavior $R_H^* \sim 1/e_0 n_h$, which has 
been in fact qualitatively reproduced using the same method
\cite{prel1}. In Fig.~3 we present the results for $R_H(n_l)$ as they
develop with the attractive potential $V>0$, already in the presence
of small stabilizing $h=0.1 t$. From results it is evident that for $V
\gg 0$ Hall constant is suppressed $|R_H|\ll R_H^*$ for 
$n_l=1/2$ but as well in a wider range of stripe filling $n_l \sim
1/2$.  On the other hand, at lowest doping $n_h=1/N$ we find large
$|R_H| \gg 1$ (even changing the sign) which is an indication that the
system is close to an insulator with small $D_{xx} \propto
E^0_{\theta\theta}$ and consequently very sensitive $R_H$ due to
Eq.(\ref{eq5}).  Results in Fig.~4 show the effect of $h$ on $R_H$ for
fixed $n_l=1/2$. We observe a systematic decrease of $R_H$ with $h$,
and consequently $R_H \sim 0 \ll R_H^*$ even at quite modest $V/t$.

\begin{figure}[t]
\begin{center}
\epsfig{file=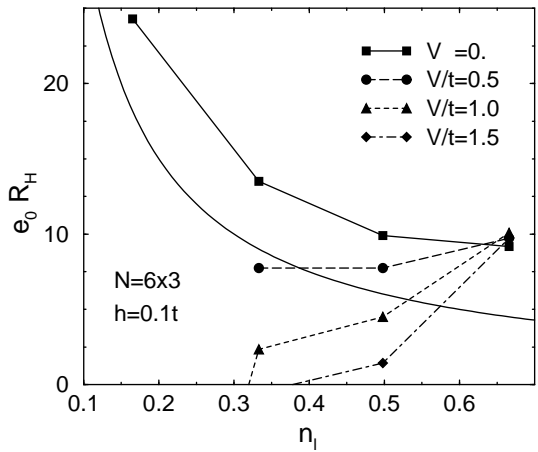,height=70mm,angle=-90}
\end{center}
\caption{Hall constant $e_0 R_H$ vs.\ stripe filling $n_l$ for a
$6\times 3$ ladder system for fixed $h=0.1t$ and various $V/t$.
Solid curve represents the semi-classical result $R_H^*=1/e_0 n_h$.}
\label{fig3}
\end{figure}

Similar conclusions can be reached following $D(n_l)$ for the same
$6\times 3$ system, as presented in Fig.~5 for $h=0.4~t$. $D$ develops
from quite monotonously increasing function of doping (with some
finite-size even-odd effect in $N_h$) at $V\sim 0, h\sim 0$ to a
dependence with a broad maximum around $n_l \sim 1/2$ for larger $V$
and $h$, as seen in Fig.~5, consistent with the vanishing of $R_H$ through
Eq.(\ref{eq1}). Such a broad maximum is also in accord with results on
the $4\times 4$ system in Fig.~1.

Let us return in conclusion to the relevance of above results to the
physics of LNSCO.  Our calculations confirm that stripe structures
with filling $n_l=1/2$ indeed have $R_H \sim 0$, due to a maximum
$D_{xx}$ closely related to minimum hole kinetic energy in such
configurations. Such a result should be directly relevant to the
commensurate case, e.g., $x=1/8$ with the inter-stripe distance
$d_0=4$ in LNSCO.  In our study we however evaluate only $R_H$ and
$D(n_h)$ imposing static stripes corresponding to fixed $d_0$ between
stripes, so we can only conjecture the behavior in the regime around
the commensurate values of $x$. Going beyond $x>1/8$ in LNSCO the
stripes are not static (and finally disappear above the LTT-LTO
transition) so our assumption of the broken symmetry is not justified
any more. One can argue that the loss of the (nearly) long range
stripe order leads to the situation closer to normal phase, i.e., to an
increasing $D(n_h)$ and hence to $R_H>0$ approaching the behavior $R_H
\sim R_H^*$ in the reference LSCO. In fact already our results in
Fig.~3 indicate that $R_H>0$ for $n_l>1/2$. On the other hand, for
$x<1/8$ experiments reveal quite static stripes with an incommensurate
spin density modulation corresponding to $d_0 = 2/x$.  Following our
results and arguments at fixed $d_0$ and for the spin ordering
induced by the staggered field $h$ we would again expect maximum
$D(n_h)$ and consequently $R_H \sim 0$ at the specific stripe filling
$n_l=1/2$. Therefore one can speculate that $R_H \sim 0$ persists in
the whole regime of static stripes. This would be definitely a very
peculiar and striking novel phenomenon, to be resolved by experiments
which have so far not been performed in LNSCO well below $x<0.12$.

\begin{figure}[t]
\begin{center}
\epsfig{file=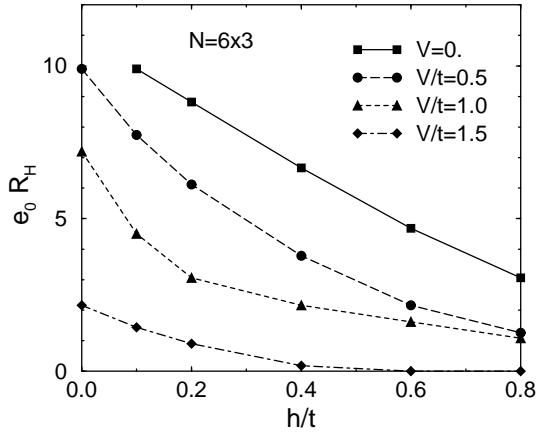,height=70mm,angle=-90}
\end{center}
\caption{$e_0R_H$ vs. $h/t$ for a $6\times 3$ ladder system
for fixed stripe filling $n_l=1/2$ and various $V/t$.}
\label{fig4}
\end{figure}
\begin{figure}[t]
\begin{center}
\epsfig{file=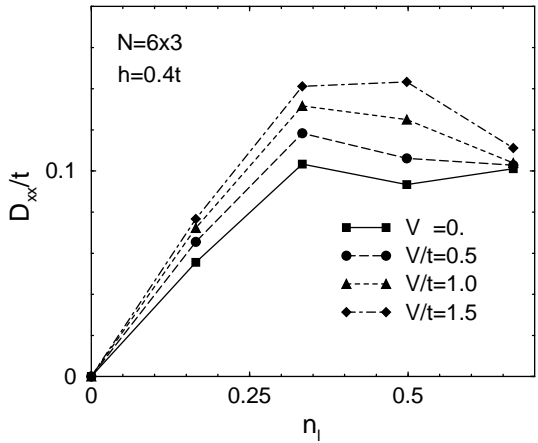,height=70mm,angle=-90}
\end{center}
\caption{$D_{xx}/t$ vs. $n_l$ for a $6\times 3$ ladder system
for fixed $h/t=0.4$ and various $V/t$.}
\label{fig5}
\end{figure}

One of the authors (P.P.) acknowledges the support of the Japan
Society for the Promotion of Science and the hospitality of the IMR,
Tohoku University, Sendai, where this work has been initiated. The
authors wish to thank also N.P. Ong, S. Uchida, J. Zaanen and X. Zotos
for helpful discussions.  This work has been supported in part by
Grand-in-Aid for Scientific Research on Priority Areas from the
Ministry of Education, Culture, Sports, Science and Technology of
Japan, NEDO and CREST, and by Ministry of Science and Technology of
Slovenia. The numerical calculations were performed in Supercomputer
Center in ISSP, University of Tokyo, and Supercomputing facilities in
IMR, Tohoku University.

\end{document}